\documentclass[12pt,preprint]{aastex}

\shorttitle{Outflow and disk in V Hya}
\shortauthors{Hirano et al.}

\received{2004 May 1}
\begin{document}

\title{HIGH VELOCITY BIPOLAR OUTFLOW AND DISK-LIKE ENVELOPE IN THE CARBON STAR V HYA}

\author{Naomi HIRANO\altaffilmark{1}\altaffilmark{2}, Hiroko 
SHINNAGA\altaffilmark{3}, DINH-V-TRUNG\altaffilmark{1},
David FONG\altaffilmark{3}, Eric KETO\altaffilmark{4},
Nimesh PATEL\altaffilmark{4}, Chunhua QI\altaffilmark{4},
Ken YOUNG\altaffilmark{4}, 
 Qizhou ZHANG\altaffilmark{4}, 
\& Junhui ZHAO\altaffilmark{4}}

\altaffiltext{1}{Academia Sinica,
Institute of Astronomy \& Astrophysics,
P.O. Box 23--141, Taipei, 106, TAIWAN, R.O.C.}
\altaffiltext{2}{e-mail: hirano@asiaa.sinica.edu.tw}

\altaffiltext{3}{Harvard-Smithsonian Center for Astrophysics,
645 North A'ohoku Place, Hilo HI 96720, U.S.A.}

\altaffiltext{4}{Harvard-Smithsonian Center for Astrophysics, 60 Garden
Street, Cambridge, MA 02138, U.S.A.}

\begin{abstract}
Using the partially completed Submillimeter Array with five 
antennas, we have observed the CO $J$=2--1 and 3--2 emission from the envelope surrounding the carbon star V Hya. 
The high-angular resolution (2$''$--4$''$) maps show that V Hya is powering a bipolar molecular jet having an extreme velocity of 70--185 km s$^{-1}$.
The axis of this high velocity jet is perpendicular to the major axis of the flattened disk-like envelope, which is expanding with a velocity of $\sim$16 km s$^{-1}$.
There is a third kinematic component, a medium-velocity wind having a deprojected velocity of 40--120 km s$^{-1}$ moving along the disk plane.
Both the high velocity jet and the medium velocity wind have a dynamical time scale of a few hundred years. 
The flattened structure and the collimated jet observed in V Hya suggests that the formation of asymmetrical structure proceeds while the central star is still in the AGB phase.

\end{abstract}


\section{INTRODUCTION}

V Hya, an evolved star located at 380 pc away from the sun \citep{Kna97}, 
is a carbon star with a high mass loss. 
Although the optical properties of V Hya are those of the normal N type carbon star, the envelope surrounding this star shows several peculiar properties.
Single-dish molecular line observations have shown that
the envelope of V Hya is elongated along the north-south direction with a velocity gradient along its minor axis \citep[e.g.,][]{Kah96}. 
Such morphological and kinematical properties are different from 
those of the typical asymptotic giant branch (AGB) envelope, which has spherically symmetric shape and uniform expansion. 
In addition, V Hya is associated with a very fast ($>$ 100 km s$^{-1}$) outflow that was observed in the optical and infrared spectra \citep[e.g.,][]{Sah88, Llo91}, and in the CO $J$=2--1 and 3--2 spectra \citep{Kna97}.
\citet{Kna97} also found that the high-velocity CO gas is bipolar, 
and proposed that the fast-moving gas is expanding along the east-west direction, perpendicular to the major axis of the envelope.
However, previous observations did not bring us the detailed structure of the high-velocity outflow and the relation 
between the high-velocity flow and the slowly expanding envelope.

\section{OBSERVATIONS}

The observations were carried out between 
February 2003 and May 2003,
with the five antennas of the partially completed Submillimeter Array
(SMA)\altaffilmark{4} at Mauna Kea, Hawaii \citep{Ho04}.
The primary-beam size (HPBW) of the 6 m antennas at 230 and 345 GHz were measured to be $\sim$54$''$ and $\sim$36$''$, respectively.
We obtained the data in two array configurations, which provided 15
independent baselines.
The spectral correlator had a bandwidth of 656 MHz
and a frequency resolution of 812.5 kHz.

The visibility data were calibrated using the MIR package \citep{Sco93}. 
We used 1055+018 or 3C273 as a phase and amplitude calibrator, and Callisto as a flux calibrator.
The bandpass was calibrated by observations of Jupiter or Saturn. 
The calibrated visibility data were Fourier transformed and CLEANed
by using the Astronomical Image Processing System (AIPS) with natural
weighting.
The synthesized beam had a size of 4.4$''$${\times}$3.1$''$ with a position angle of $-$21$^{\circ}$ at 230 GHz and 2.1$''$${\times}$1.9$''$ with a position angle of $-$28$^{\circ}$ at 345 GHz.
In making maps, we averaged 2 channels in CO $J$=2--1 and 3 channels
in CO $J$=3--2 and made maps with velocity resolution of 2.1 km s$^{-1}$.
The rms noise level of the maps at the velocity resolution of 2.1 km s$^{-1}$ was 0.09 Jy beam$^{-1}$ (0.15 K in $T_{\rm B}$)
in CO $J$=2--1 and 0.3 Jy beam$^{-1}$ (0.77 K in $T_{\rm B}$) in CO $J$=3--2.
We averaged the entire 656 MHz of both image sidebands
to make the maps of 241 GHz and 335 GHz continuum emission.
Continuum maps were made with uniform weighting, which provided
a synthesized beam of 3.0$''$$\times$2.2$''$ with a position angle 
of 37$^{\circ}$ at 241 GHz and 2.0$''$$\times$1.3$''$ with
a position angle of 48$^{\circ}$ at 335 GHz.
The rms noise levels of the 241 GHz and 335 GHz continuum maps were 
9 mJy beam$^{-1}$ and 20 mJy beam$^{-1}$, respectively.

\altaffiltext{'S}{The Submillimeter Array is a joint project between the Smithsonian Astrophysical Observatory and the Academia Sinica Institute of Astronomy and Astrophysics, and is funded by the Smithsonian Institution and the Academia Sinica.}

\section{RESULTS AND DISCUSSION}

\subsection{Continuum emission from V Hya}

We detected the continuum emission
peaks at 
${\alpha}_{J2000}$ = 10$^{\rm h}$ 51$^{\rm m}$ 37.25$^{\rm s}$,
${\delta }_{J2000}$ = $-$21$^{\circ} $15$'$ 00.5$''$
at both 241 and 335 GHz. 
The position of the continuum source coincides well with the stellar position at the observational epoch (J2003.25), which is  derived from the coordinates and proper motion measured by the Hipparcos satellite \citep{Per97}.
Figures 1k and 1l show that the continuum source is not resolved with our beams.
This suggests that the most of the emission comes from a spatially compact region with a radius of $\lesssim$1$''$ (380 AU).
The flux density of the source is 63$\pm$16 mJy at 241 GHz and 105$\pm$39 mJy at 335 GHz.
The spectral index $\alpha$ between 241 GHz and 335 GHz is $\sim$1.6, which is shallower than Rayleigh-Jeans. 
This might imply the contribution of the free-free emission from shock-ionized gas close to the star \citep{Lut92, Kna97}.
However,  because of the large uncertainty in the flux at 335 GHz, it is difficult to estimate the contribution of the ionized component to the submillimeter spectra.
If we assume that the continuum emission at 241 and 335 GHz arises from the thermal emission from the dusty envelope, 
the mass of dust $M_{dust}$ required to produce the observed flux densities is estimated by
\begin{equation}
M_{dust}{\ge}{{2a\rho _{dust}D\lambda ^2(S_\nu-S_*) } \over {3{\cal Q}_{\nu}kT_{dust}}},
\end{equation}
in the Rayleigh-Jeans limit.
Here, $S_{\nu}$ is the observed flux, $S_{*}$ is the flux that comes from the stellar photosphere, and D is the source distance.
At 241 GHz, $S_{*}$ is estimated to be $\sim$10 mJy.
We assume that the dust particles have radius $a$ of 2000 {\AA}, material density $\rho_{dust}$ of 2.25 g cm$^{-3}$, and
emissivity ${\cal Q}_{\nu}$ of 6.17$\times$10$^{-4}$ at 274.6 GHz with a power-law dependence on frequency with index $\beta$ = 1.
We adopted a grain temperature to be 360 K.
This corresponds to the grain temperature at $r=$380 AU if it has a power-law dependence on radius, $T_{g}=T_{*}(R_{*}/r)^{2/(4+{\beta})}$ \citep{Kna93}, with an effective temperature $T_{*}$ and a radius $R_{*}$ of the star of 2650 K and 3.8$\times$10$^{13}$ cm \citep{Kna99}, respectively.
The mass of dust was calculated to be 6.5$\times$10$^{-6}$ $M_{\odot}$. 
If we assume a constant mass-loss, the dust mass loss rate is estimated to be 5.7$\times$10$^{-8}$$M_{\odot}$ yr$^{-1}$ based on an outflow speed of $\sim$16 km s$^{-1}$ (see section 3.4).

\subsection{Flattened envelope and high-velocity bipolar outflow}

Strong CO $J$=2--1 and $J$=3--2 emission was detected in the
velocity range from $V_{\rm LSR}=-$45 km s$^{-1}$ to +12 km s$^{-1}$, 
corresponding to $\pm$$\sim$28 km s$^{-1}$ from the systemic velocity, $V_{\rm sys}=$$-$17.5 km s$^{-1}$.
Both CO $J$=2--1 and 3--2 maps show that the molecular envelope is elongated along the north-south direction.
We have smoothed our interferometric maps to mimic the CO $J$=2--1 and 3--2 spectra observed with the 30$''$ beam and 20$''$ beam, respectively, and compared the spectra observed by \citet{Kna97} and \citet{Sta95}.
We found that $\sim$50 \% of the CO $J$=2--1 flux and $\sim$35 \% of the CO $J$=3--2 flux observed by the single-dish telescope was recovered by the SMA.
 
The velocity-channel maps of CO $J$=3--2 emission are shown in Figures 1a to 1i.
The velocity structure traced by the CO $J$=2--1 emission is similar to that shown in the CO $J$=3--2 maps.
We found that the approaching part of the envelope is located to the west and the receding part to the east of the star,
which is consistent with the previous single-dish CO $J$=2--1 results of \citet{Kah96}. 
The spatial-kinematic structure shown in Figure 1 can be explained if the CO emission arises from an inclined flattened envelope expanding along its radial direction.
A flattened disk-like structure was also identified by \citet{Sah03} based on their interferometric CO $J$=1--0 observations.
However, the observed velocity pattern does not completely match that of a uniformly expanding disk.
The largest spatial extents of $\pm$7$''$ toward the east and west are shown in the velocity channels of $\pm$$\sim$8 km s$^{-1}$ from $V_{\rm sys}$ (Figures 1g and 1c, respectively), and not in the highest velocity channels.
On the other hand, the higher velocity CO emission shown in Figures 1a and 1i comes from the regions  close to the central star.
The position-velocity diagram of CO $J$=2--1 along the east-west cut presented in Figure 2b suggests that there are two kinematic components, that is, 
two vertical ridges at ${\Delta}V{\sim}$$\pm$8 km s$^{-1}$ from $V_{\rm sys}$ and a horizontal ridge that extends to $\pm$30 km s$^{-1}$.
The $\pm$8 km s$^{-1}$ components correspond to the two spikes in the CO line profiles observed by \citet{Kah96} and \citet{Kna97}, and are likely to arise from the expanding disk inclined to the line of sight (hereafter referred to as the low-velocity disk).
The $\pm$30 km s$^{-1}$ component has a clear bipolarity with the redshifted part to the east and the blueshifted part to the west (referred to as the medium-velocity wind). 


Both CO $J$=2--1 and 3--2 lines are associated with the broad wing emission that extends to $\pm$$\sim$150 km s$^{-1}$ from $V_{\rm sys}$.
The detection of these wing components is confident because the phases  of the visibilities show coherent behavior across the velocity range of $\pm$150 km s$^{-1}$.
In Figures 2a and 2c, we show the position-velocity diagrams of the CO $J$=2--1 along the east-west cut in the blueshifted (${\Delta}V$ from $V_{\rm sys}$ $<$ $-$40 km s$^{-1}$) and redshifted (${\Delta}V$ $>$ +40 km s$^{-1}$) velocity ranges, respectively. 
In order to improve the signal to noise ratio, we made the maps of these velocity ranges by averaging 10 channels, providing the effective velocity resolution of 10.6 km s$^{-1}$.
In order to exhibit the features clearly, we smoothed the maps to 5$''$ before we create these three position-velocity diagrams.

Figures 2a and 2c suggest that emission in the velocity ranges of ${\Delta}V < \pm$60 km s$^{-1}$ corresponds to the lower intensity part of the medium-velocity wind.
On the other hand, the velocity structure of the higher velocity component with ${\Delta}V > \pm$60 km s$^{-1}$ is different. 
The spatial distribution of the high-velocity component (${\Delta}V$ from $V_{\rm sys}$ is $\pm$(60--162) km s$^{-1}$) shown in Figure 3 superposed on the intensity-weighted mean velocity map of CO $J$=2--1.
The high-velocity component has its approaching part in the east and the receding component in the west of the star.
This orientation is {\it opposite} with respect to the lower velocity components as suggested by \citet{Kna97}.
Such an orientation can be explained if the high velocity component is a pair of collimated jets blown out in the polar direction of the low-velocity disk. 
Hereafter, we refer to this component as the high-velocity jet.
Recently, \citet{Sah03} discovered the [S II] emission-line blob with high negative radial velocity of ${\Delta}V{\sim}$$-$240 km s$^{-1}$ at 0.2$''$ east of the central star.
This also suggests the presence of the collimated high velocity jet with approaching part to the east of the central star.
In the CO $J$=2--1, the receding part of the bipolar jet, which has not observed in the [S II] emission probably due to the absorption caused by the obscuring disk, is clearly detected in the west of the star.
Figures 2a and 2c show that the negative and positive velocities of the high velocity jet increase as the positional offset from the star increases.
Such a velocity increase can be explained if we take into account that the density of the circumstellar envelope decreases rapidly along the east-west direction.
Figures 2a, 2c and 3b show that the high velocity component consists of discrete clumps separated by $\sim$5$''$, suggesting that the high velocity jet is episodic.

\subsection{Circumstellar structure of V Hya}

The high-angular resolution CO $J$=3--2 and 2--1 maps
suggest that the circumstellar structure of V Hya consists of three kinematic components; the low-velocity disk having a radial velocity offsets of ${\Delta}V$ = $\pm$8 km s$^{-1}$ from $V_{\rm sys}$ , the medium-velocity wind with ${\Delta}V<{\pm}$60 km s$^{-1}$, and the high-velocity jet with ${\Delta}V$=$\pm$60--160 km s$^{-1}$. 
These three components corresponds to the $\pm$8 km s$^{-1}$ spikes, the 45 km s$^{-1}$ wind, and the 200 km s$^{-1}$ wind defined by \citet{Kna97}, respectively.

The spatial extent of the low-velocity disk along the major (N-S) and minor (E-W) axes measured at the 3 $\sigma$ level are $\pm$8$''$ and $\pm$7$''$, respectively in the CO $J$=3--2 and $\pm$15$''$ and $\pm$13$''$, respectively in the CO $J$=2--1.
If we assume that the disk is thin, the ratio between the major and minor axes suggests that the disk plane is inclined by $\sim$60$^{\circ}$ from the line of sight, which is consistent with the the previous estimation of $\sim$70$^{\circ}$ by \citet{Sah03}.
We find that the gas in the low-velocity disk is moving at almost constant deprojected velocity of 16 km s$^{-1}$, which is similar to the typical AGB wind velocity.
The north-south extent in the CO $J$=2--1 map suggests that the low-velocity disk has a radius of 5700 AU and a dynamical time scale of 1700 yr.
If we assume that the axis of the high velocity jet is perpendicular to the disk plane, the deprojected velocity and the dynamical time scale of the jet are estimated to be 70--185 km s$^{-1}$ and 100--250 yr, respectively.

What is the nature of the  medium-velocity wind?
Here we present three possible scenarios;  1) the interaction between the low-velocity disk and the high-velocity wind, 2) equatorial mass outflow with higher velocity, or 3) a second bipolar outflow with different axis.
The first case assumes that the dense equatorial disk collimates the spherically expanding or poorly collimated high-velocity wind into the polar direction. 
In this case, the wind directed to the polar direction can escape rather freely, while the one directed to the equatorial plane pushes the dense molecular gas in the disk plane, and produces the medium-velocity wind.
The second case is possible if the velocity of the recent mass flow along the equatorial disk plane has been remarkably enhanced for some reason.
In either case 1 or case 2, the direction of the medium-velocity wind is close to the disk plane.
Therefore, the deprojected velocity of the wind is estimated to be 40--120 km s$^{-1}$.
The dynamical time scale of the medium-velocity wind estimated from the spatial extent of $\sim$5--8$''$, inclination angle, and the deprojected velocity is 100--400 yr, which is close to the time scale of the high-velocity jet.
The third case is the interpretation proposed by \citet{Kah96}, in which the medium-velocity wind came from a bipolar outflow with large opening angle.
However, the orientation of the approaching and receding parts of the medium-velocity wind suggests that the axis of this bipolar flow should be close to the plane of the disk.
If we take into account the geometry of the low-velocity disk and the high-velocity jet, the case 3 scenario is less likely, although we cannot rule out the possibility that V Hya ejects two orthogonal outflows as in the case of CRL 2688 \citep[e.g.,][]{Cox00}.
In Figure 4, we show a schematic picture of these three kinematic components based on the scenarios of case 1 or 2. 

The origin of the disk and jet system like V Hya has not yet been elucidated. 
However,  theoretical models \citep[e.g.,][]{Sok97, Sok00} suggest that it is difficult for a single star to form a flattened disk-like structure.
The evidence of rapid rotation of V Hya \citep{Bar95} suggests that this star has a companion.
It is likely that the close companion plays an important role in enhancing the equatorial mass loss, and producing the disk and jet structure \citep [e.g.,][]{Sok92, Mor87}.
The circumstellar structure of V Hya resembles that of the proto-PN, CRL 618, which is also associated with large equatorial tori and extremely high-velocity bipolar outflow \citep{San04}.
The circumstellar structure of V Hya implies that the asymmetric structure with disk and jet appears when the central star is still in the AGB phase.

\acknowledgments

We wish to thank all the SMA staff in Hawaii, Cambridge, and Taipei
for their enthusiastic help during these observations.
We also thank P.T.P. Ho, J.M. Moran, and anonymous referee for helpful comments on the manuscript.

\clearpage

\begin{figure}
\epsscale{0.75}
\plotone{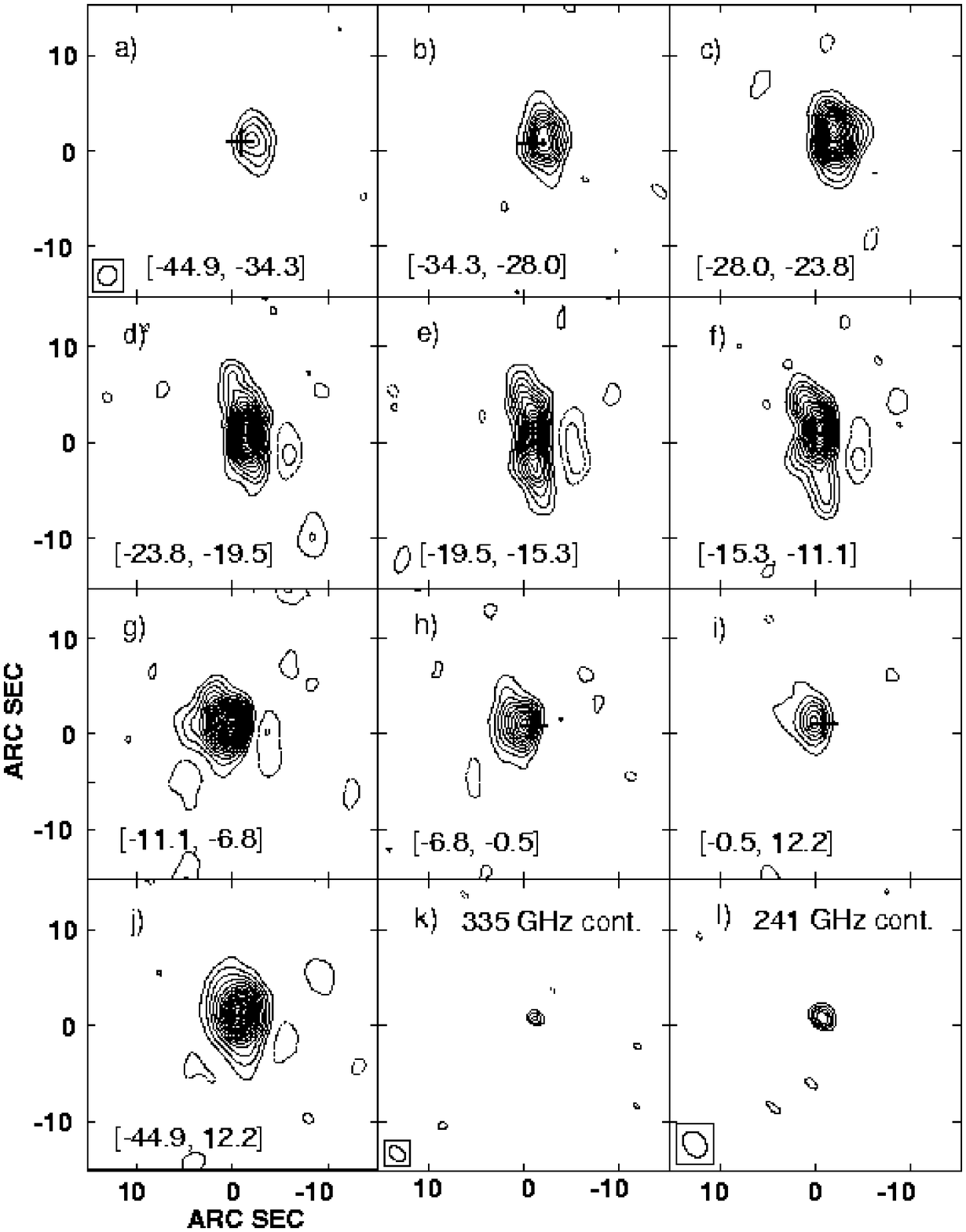}
\caption{(a)--(i) Velocity-channel maps of the CO $J$=3--2 line.
The coordinates of the map center are ${\alpha}_{J2000}$ = 
10$^{\rm h}$ 51$^{\rm m}$ 37.25$^{\rm s}$, ${\delta }_{J2000}$ = 
-21$^{\circ} $15$'$ 00.5$''$. 
The velocity range of each panel is shown in a bracket.
Cross denotes the position of the continuum source (see panels k and l).
The contours are drawn every 3$\sigma$.
The 3$\sigma$ level of each map is 4.2 Jy beam$^{-1}$ km s$^{-1}$ in (a), 3.3 Jy beam$^{-1}$ km s$^{-1}$ in (b) and (h), 2.7 Jy beam$^{-1}$ km s$^{-1}$ in (c)--(g), and 4.7 Jy beam$^{-1}$ km s$^{-1}$ in (i).
(j) A map of CO $J$=3--2 integrated over the velocity range from $-$44.9 to +12.2 km s$^{-1}$. 
The contours are drawn every 3$\sigma$ (13.7 Jy beam$^{-1}$ km s$^{-1}$) step.
(k) A map of the 335 GHz continuum emission.
The lowest contours are 60 mJy beam$^{-1}$ (3$\sigma$) with the step of 20 mJy beam$^{-1}$ (1$\sigma$).
(l) A map of the 241 GHz continuum emission .
The contours are every 9 mJy beam$^{-1}$
(1$\sigma$) with the lowest contour of 27 mJy beam$^{-1}$ (3$\sigma$).
\label{fig1}}
\end{figure}

\clearpage 

\begin{figure}
\plotone{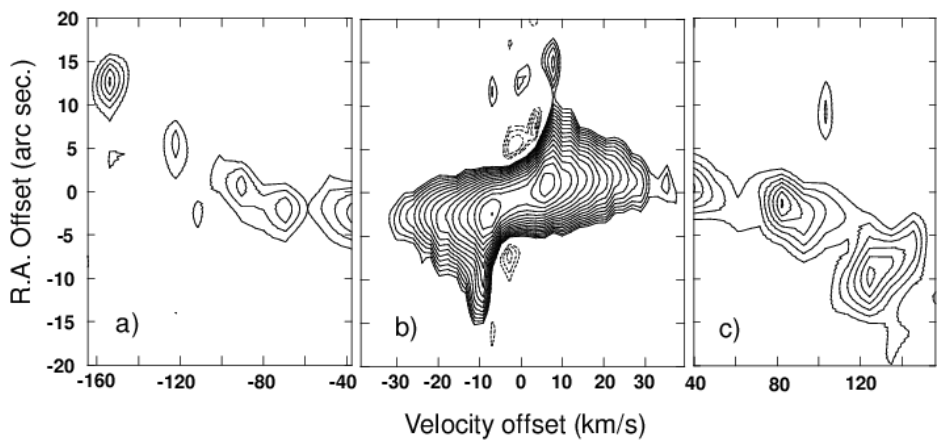}
\caption{Position-velocity map of CO $J$=2--1 along the east-west cut through the field center in the three different velocity ranges. The effective velocity and spatial resolutions are 10.6 km s$^{-1}$ and 5.0$''$, respectively in (a) and (c), and are 2.1 km s$^{-1}$ and 5.0$''$, respectively in (b). 
The contour interval is 0.055 Jy beam$^{-1}$ (1$\sigma$) with the lowest contour at 0.11 Jy beam$^{-1}$ (2$\sigma$) in (a) and (c).
The contours in (b) are drawn in a log scale; 0.40, 0.50, 0.63, 0.80, 1.00, 1.3, 1.6, 2.0, 2.5, 3.2, 4.0, 10.0, 12.6, 15.9 Jy beam$^{-1}$.
\label{fig2}}
\end{figure}

\clearpage 

\begin{figure}
\plotone{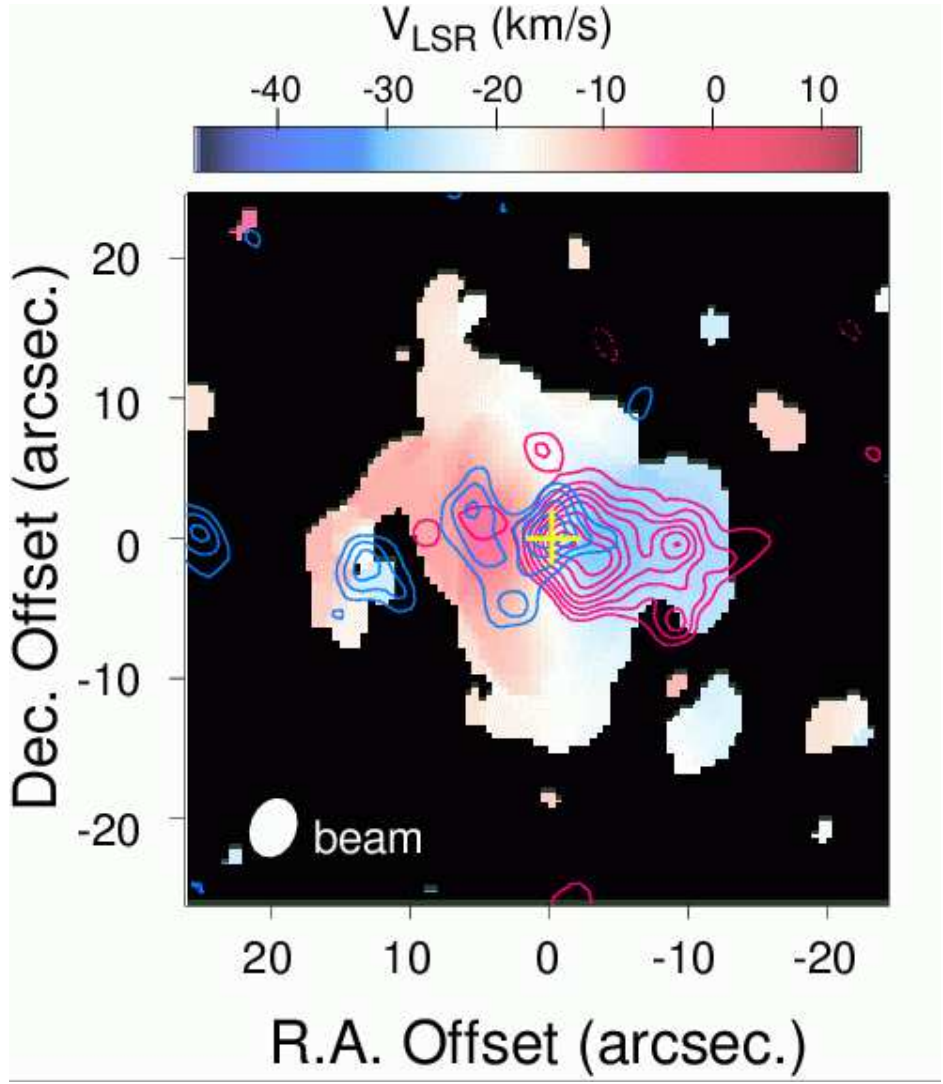}
\caption{The CO $J$=2--1 maps integrated over the velocity range of ${\pm}$(60.0--162.2) km s$^{-1}$ from the systemic velocity of $-$17.5 km s$^{-1}$ (blue and red contours) superposed on the intensity-weighted mean velocity map of the CO $J$=2--1 in the velocity range of $\pm$30 km s$^{-1}$ from the systemic velocity. 
The contours are drawn every 14.3 Jy beam$^{-1}$ km s$^{-1}$ (1$\sigma$)
with the lowest contours of 42.9 Jy beam$^{-1}$ km s$^{-1}$ (3$\sigma$).
\label{fig3}}
\end{figure}

\clearpage 

\begin{figure}
\plotone{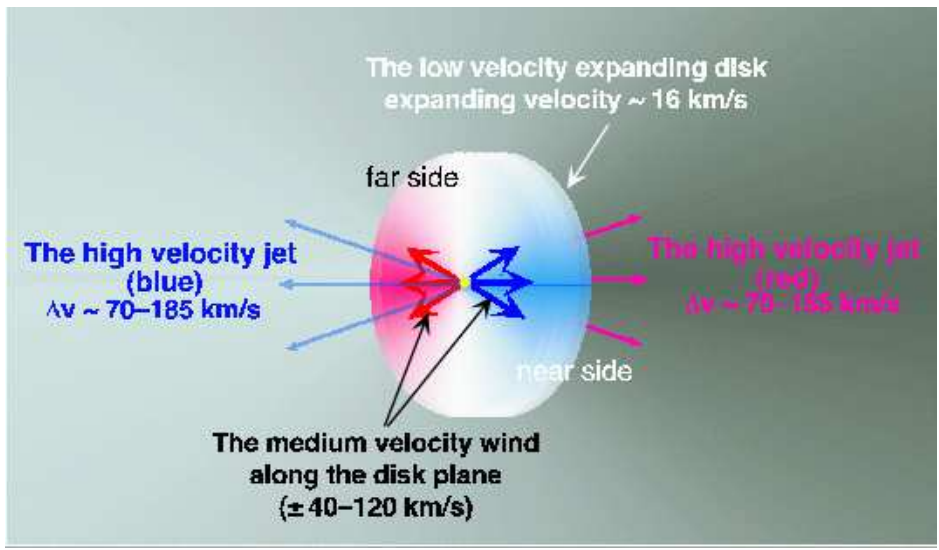}
\caption{Schematic picture of the circumstellar structure of V Hya.
 \label{fig4}}
\end{figure}

\end{document}